\documentstyle[12pt]{article}

\begin{document}
\setcounter{page}{1}
\renewcommand{\thefootnote}{\fnsymbol{footnote}}
\pagestyle{plain} \vspace{1cm}
\begin{center}
\Large{\bf  Hawking Radiation as Quantum Tunneling from
Noncommutative Schwarzschild Black Hole } \vspace{1.5cm}

\small \vspace{1cm} {\bf Kourosh Nozari$^{\rm a,}$\footnote{
knozari@umz.ac.ir}}\quad\quad and \quad\quad {\bf S. Hamid
Mehdipour$^{\rm
a,b,}$\footnote{h.mehdipour@umz.ac.ir}} \\
\vspace{0.5cm} {\it $^{a}$Department of Physics, Faculty of Basic
Sciences,
University of Mazandaran,\\
P. O. Box 47416-1467, Babolsar, IRAN}\\
and\\

{\it $^{b}$ Islamic Azad University, Lahijan Branch,\\
P. O. Box 1616, Lahijan, IRAN }\\

\end{center}
\vspace{1.5cm}
\begin{abstract}
We study tunneling process through quantum horizon of a
Schwarzschild black hole in noncommutative spacetime. This is done
by considering the effect of smearing of the particle mass as a
Gaussian profile in flat spacetime. We show that even in this
noncommutative setup there will be no correlation between the
different modes of radiation which reflects the fact that
information doesn't come out continuously during the evaporation
process at least at late-time. However, due to spacetime
noncommutativity, information might be preserved by
a stable black hole remnant.
\end{abstract}
\newpage

\section{ Introduction}
In 1975, Hawking proposed a scenario in which black hole can radiate
from its event horizon as a black body with a purely thermal
spectrum at the temperature $T_H=\frac{\hbar c^3\kappa}{2\pi
k_{B}G}$, utilizing the procedure of quantum field theory in curved
spacetime ($\kappa$ is the surface gravity that demonstrates the
strength of the gravitational field near the black hole surface).
This leads us to a non-unitary quantum evolution where maps a pure
state to a mixed state. In 2000, Parikh and Wilczek [2] proposed the
method of null-geodesic to derive Hawking temperature as a quantum
tunneling process. In this quantum tunneling framework, the form of
the corrected radiation is not exactly thermal which yields a
unitary quantum evolution. However, their form of the correction for
emission is not adequate by itself to retrieve information since it
fails to find the correlations between the emission rates of
different modes in the black hole radiation spectrum. Possibly,
spacetime noncommutativity [3-5], that is, an inherent trait of the
manifold by itself and the fact that spacetime points might be
noncommutative, opens the way to find a solution to the {\it black
hole information paradox} that can be solved by ceasing the black
hole to decay beyond a {\it minimal mass} $M_0$. In 2003, Smailagic
and Spallucci [6-8] postulated a new attractive model of
noncommutativity in terms of coherent states which satisfies Lorentz
invariance, Unitarity and UV-finiteness of quantum field theory. In
2005, Nicolini, Smailagic and Spallucci (NSS) [9] by using this
method have found the generalized line element of Schwarzschild
spacetime based on coordinate coherent state noncommutativity. It
has been shown that the generalized line element does not permit the
black hole to decay lower than $M_0$. Thus, the evaporation process
finishes when black hole approaches a Planck size remnant with zero
temperature, which does not diverge at all, rather it reaches a
maximum value before shrinking to absolute zero. Since spacetime
noncommutativity can eliminate some kind of divergences (which
appear in General Relativity), and also is an intrinsic property of
the manifold itself (even in the absence of gravity), we hope to
cure a step further and modify the tunneling paradigm utilizing the
noncommutative field theory. In this manner, we would like to
proceed the Parikh-Wilczek tunneling process using a fascinating
formulation of noncommutativity of coordinates that is carried out
by the Gaussian distribution of coherent states.

\section{Noncommutative Schwarzschild Black Hole}
A valuable test of spacetime noncommutativity is its possibly
observable effects on the properties of black holes. To inquire into
this issue, one would require to prosperously build the
noncommutativity corresponding to the General Relativity. Although
this issue has been considered in the literature [10], but no
perfect and wholly convincing theory of this model yet exists. There
are plenty formulations of noncommutative field theory established
upon the Weyl-Wigner-Moyal $\star$-product [11] that conduct to
downfall in finding a solution to the some prominent difficulties,
such as Lorentz invariance breaking, defeat of unitarity and UV
divergences of quantum field theory. The incident of
noncommutativity at a observable scale has the ability which leads
to important effects in the expected properties of the black holes.
Although a perfect noncommutative theory of gravity does not yet
exist, it becomes essential to model the noncommutativity effects in
the frame of the commutative General Relativity. Lately, the authors
in Ref. [6-8] have regarded a physically inspired and obedient type
of the noncommutativity amendments to Schwarzschild black hole
solutions (coordinate coherent states formalism), that can be
released from the difficulties mentioned above. In this formalism,
General Relativity in its common commutative case as characterized
by the Einstein-Hilbert action stays appropriate. If
noncommutativity effects can be behaved in a perturbative manner,
then this comes into view defensible, at least to a good
approximation. The authors in Ref. [10] have really demonstrated
that the leading noncommutativity amendments to the form of the
Einstein-Hilbert action are at least second order in the
noncommutativity parameter, $\theta$. The generalization of quantum
field theory by noncommutativity based on coordinate coherent state
formalism is also interestingly curing the short distance behavior
of pointlike structures [6-9] (see also [12]). In this approach, the
particle mass $M$, instead of being completely localized at a point,
is dispensed throughout a region of linear size $\sqrt{\theta}$,
that the implementation of these arguments leads to substitution of
position Dirac-delta function, describing pointlike structures, with
Gaussian function, describing smeared structures. On the other hand,
the mass density of a static, spherically symmetric, particle-like
gravitational source cannot be a delta function distribution but
will be given by a Gaussian distribution of minimal width
$\sqrt{\theta}$ as follows
\begin{equation}
\rho_{\theta}(r)=\frac{M}{(4\pi\theta)^{\frac{3}{2}}}\exp\bigg(-\frac{r^2}{4\theta}\bigg).
\end{equation}
The Schwarzschild solution of the Einstein equations associated with
these smeared mass Gaussian function sources leads to line element
as
\begin{equation}
ds^2=-\bigg(1-\frac{2M_\theta}{r}\bigg)dt^2+
\bigg(1-\frac{2M_\theta}{r}\bigg)^{-1}dr^2+r^2 d\Omega^2,
\end{equation}
where the smeared mass distribution is implicity given in terms of
the lower incomplete Gamma function as
\begin{equation}
M_\theta=\int_0^r\rho_{\theta}(r)4\pi
r^2dr=\frac{2M}{\sqrt{\pi}}\gamma\bigg(\frac{3}{2},\frac{r^2}{4\theta}\bigg)\equiv
\frac{2M}{\sqrt{\pi}}\int_0^{\frac{r^2}{4\theta}}t^{\frac{1}{2}}e^{-t}dt.
\end{equation}
(Hereafter we set the fundamental constants equal to unity; $\hbar =
c = k_B = 1$.) In the limit of \, $\theta\rightarrow0$, one recovers
the complete Gamma function $\Gamma(\frac{3}{2})$,
\begin{equation}
\lim_{\theta\rightarrow0}M_\theta= M,
\end{equation}
and the modified Schwarzschild solution reduces to the ordinary
Schwarzschild solution. The line element (2) characterizes the
geometry of a noncommutative inspired Schwarzschild black hole. The
radiating behavior of such a modified Schwarzschild black hole can
now be investigated and can easily be shown by plotting $g_{00}$ as
a function of $r$, for different values of $M$ ( hereafter, for
plotting the figures we set the value of the noncommutativity
parameter equal to unity; $\theta=1$). Fig. 1 shows that coordinate
noncommutativity leads to the existence of a minimal non-zero mass
which black hole (due to Hawking radiation and evaporation) can
shrink to it.
\begin{figure}[htp]
\begin{center}
\includegraphics{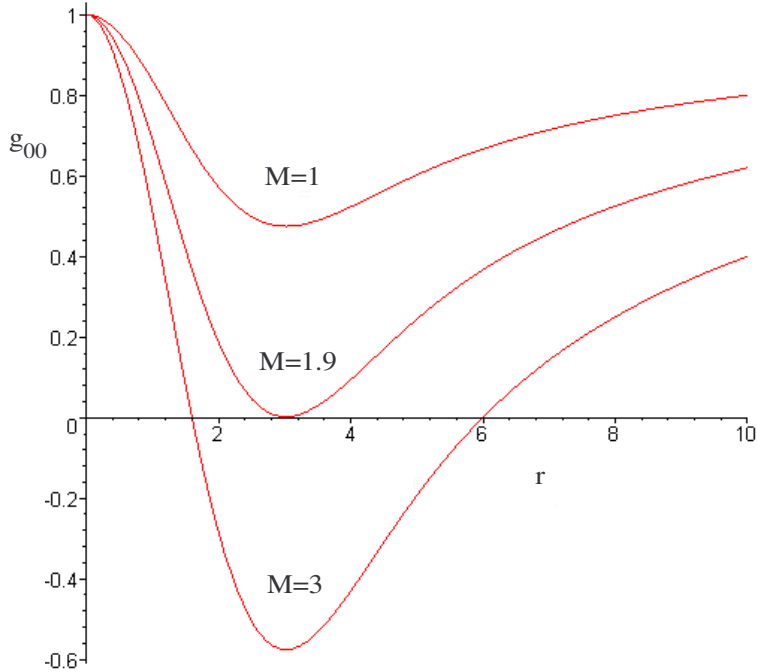}
\end{center}
\vspace{7.15 cm} \caption{\scriptsize {$g_{00}$ versus the radius
$r$ for different values of mass $M$. The figure shows the
possibility of having extremal configuration with one degenerate
event horizon when $M=M_0\approx1.9$ ({\it i.e.}, the existence of a
minimal non-zero mass), and no event horizon when the mass of the
black hole is smaller than $M_0$. Also as figure shows, the distance
between the horizons will increase by increasing the black hole mass
(two event horizons). }}
\end{figure}
The event horizon of this line element can be found where $g_{00} (
r_H ) = 0$, that is implicity written in terms of the upper
incomplete Gamma function as
\begin{equation}
r_H=2M_\theta(r_H)=2M\Bigg(1-\frac{2}{\sqrt{\pi}}\Gamma\bigg(\frac{3}{2},\frac{r_H^2}{4\theta}\bigg)\Bigg).
\end{equation}
The noncommutative Schwarzschild radius versus the mass can
approximately be calculated by setting $r_H=2M$ into the upper
incomplete Gamma function as
\begin{equation}
r_H=2M\Bigg({\cal{E}}\bigg(\frac{M}{\sqrt{\theta}}\bigg)-\frac{2M}{\sqrt{\pi
\theta}}\exp\bigg(-\frac{M^2}{\theta}\bigg)\Bigg),
\end{equation}
where ${\cal{E}}(x)$ shows the {\it Gauss Error Function} defined as
$$ {\cal{E}}(x)\equiv \frac{2}{\sqrt{\pi}}\int_{0}^{x}e^{-t^2}dt.$$
For very large masses, the ${\cal{E}}(\frac{M}{\sqrt{\theta}})$
tends to unity and second term on the right will exponentially be
reduced to zero and one retrieves
the classical Schwarzschild radius, $r_H\approx2M$.\\
When such a noncommutative black hole radiates, its temperature can
be calculated to find
\begin{equation}
T_H={1\over {4\pi}} {{dg_{00}}\over
{dr}}|_{r=r_H}=M\Bigg[\frac{{\cal{E}}\Big(\frac{r_H}{2\sqrt{\theta}}\Big)}{2\pi
r_H^2}-\frac{\exp\Big(-\frac{r_H^2}{4\theta}\Big)}{4(\pi
\theta)^{\frac{3}{2}}}\bigg(r_H+\frac{2\theta}{r_H}\bigg)\Bigg].
\end{equation}
For the commutative case,
$\frac{M}{\sqrt{\theta}}\rightarrow\infty$, one recovers the
classical Hawking temperature, $T_H=\frac{1}{8\pi M}$. The numerical
calculation of the modified Hawking temperature as a function of the
mass is presented in Fig. 2. In this modified (noncommutative)
version, not only $T_H$ does not diverge at all but also it reaches
a maximum value before dropping to absolute zero at a minimal
non-zero mass, $M=M_0\approx1.9$, which black hole shrink to it.\\
\begin{figure}[htp]
\begin{center}
\includegraphics{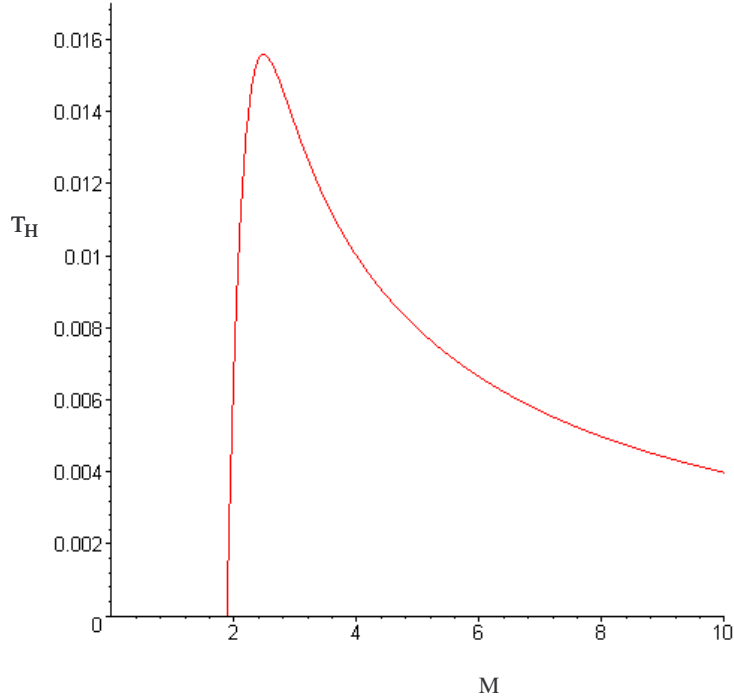}
\end{center}
\vspace{7 cm} \caption{\scriptsize { Black hole temperature, $T_H$,
as a function of $M$. The existence of a minimal non-zero mass and
disappearance of divergence are clear. }}
\end{figure}

To find the analytical form of the modified (noncommutative)
entropy, $S_{NC}$, we should note that our calculation for obtaining
the modified Hawking temperature, equation (7), is exact and no
approximation has been made. But there is no analytical solution for
entropy from the first law of classical black hole thermodynamics
$dM=T_H dS$ with $T_H$ given as (7), even if we set $r_H=2M$ in this
relation. Nevertheless, to obtain an approximate analytical form of
entropy we can use the following expression as an approximation for
noncommutative Hawking temperature
\begin{equation}
T_H=\frac{1}{4\pi r_H},
\end{equation}
where $r_H$ is given by equation (6). Eventually, the entropy of the
black hole can be obtained as analytical form using the first law,
\begin{equation}
S_{NC}=\int\frac{dM}{T_H}=2\pi\int\frac{dM}{\kappa(M)}=4\pi
M^2{\cal{E}}\bigg(\frac{M}{\sqrt{\theta}}\bigg)-6\pi\theta
{\cal{E}}\bigg(\frac{M}{\sqrt{\theta}}\bigg)+12\sqrt{\pi\theta}M\exp\bigg(-\frac{M^2}{\theta}\bigg).
\end{equation}
Where $\kappa(M)$ is the horizon noncommutative surface gravity and
is given by
\begin{equation}
\kappa(M)=\left[4M\Bigg({\cal{E}}\bigg(\frac{M}{\sqrt{\theta}}\bigg)-\frac{2M}{\sqrt{\pi
\theta}}\exp\bigg(-\frac{M^2}{\theta}\bigg)\Bigg)\right]^{-1}
\end{equation}
Behavior of the entropy $S_{NC}$, as a function of the mass is
depicted in Fig. 3. As this figure shows, at the final stage of the
black hole evaporation, the black hole ceases to radiate and its
entropy reaches zero and the existence of a minimal non-zero mass is
clear again. In the large mass limit {\it i.e.},
$\frac{M}{\sqrt{\theta}}\gg1$, one recovers the standard
Bekenstein-Hawking Entropy plus $\theta$-corrections, which leads to
\begin{equation}
S_{NC}=4\pi
M^2+12\sqrt{\pi\theta}M\exp\bigg(-\frac{M^2}{\theta}\bigg).
\end{equation}
\begin{figure}[htp]
\begin{center}
\includegraphics{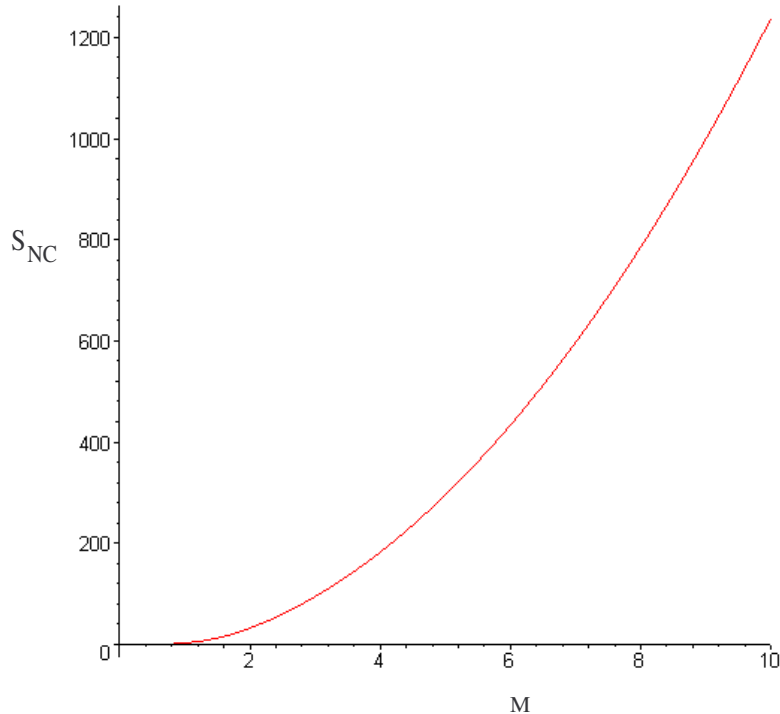}
\end{center}
\vspace{6.8cm} \caption{\scriptsize {Black hole entropy, $S_{NC}$,
as a function of $M$. Note that the figure is plotted approximately
by the equation (24). }}
\end{figure}

It should be noted that if we had picked out a different form for
the probability of matter distribution, instead of distribution (1),
solely the smeared mass distribution $M_\theta$ would be altered
however their general properties would be directed to entirely
comparable consequences to those above. For instance, we consider a
Lorentzian distribution of smeared particle
\begin{equation}
\rho_{\theta'}(r)=\frac{M\sqrt{\theta'}}{\pi^2(r^2+\theta'^2)^2}.
\end{equation}
Here the noncommutativity parameter, $\theta'$, is actually not
identical to $\theta$. The smeared mass distribution is now given by
\begin{equation}
M_{\theta'}=\int_0^r\rho_{\theta'}(r)4\pi
r^2dr=\frac{2M}{\pi}\left(\tan^{-1}\bigg(\frac{r}{\sqrt{\theta'}}\bigg)-\frac{r\sqrt{\theta'}}{(r^2+\theta')}\right).
\end{equation}
In the limit of $\theta'$ going to zero, we get
$M_{\theta'}\rightarrow M$. As expected, the smeared mass Lorentzian
distribution, $M_{\theta'}$, has the same limiting properties and is
completely comparable to the smeared mass Gaussian distribution,
$M_{\theta}$, qualitatively. Then, many of the outcomes that we
achieved stay applicable if we take the other kind of probability
distribution (see [13]). The lack of responsiveness of these
consequences to our Gaussian formalism of the smearing can easily be
exhibited by plotting $g_{00}$ as a function of $r$ for different
values of $M$ (see Fig. 4).
\begin{figure}[htp]
\begin{center}
\includegraphics{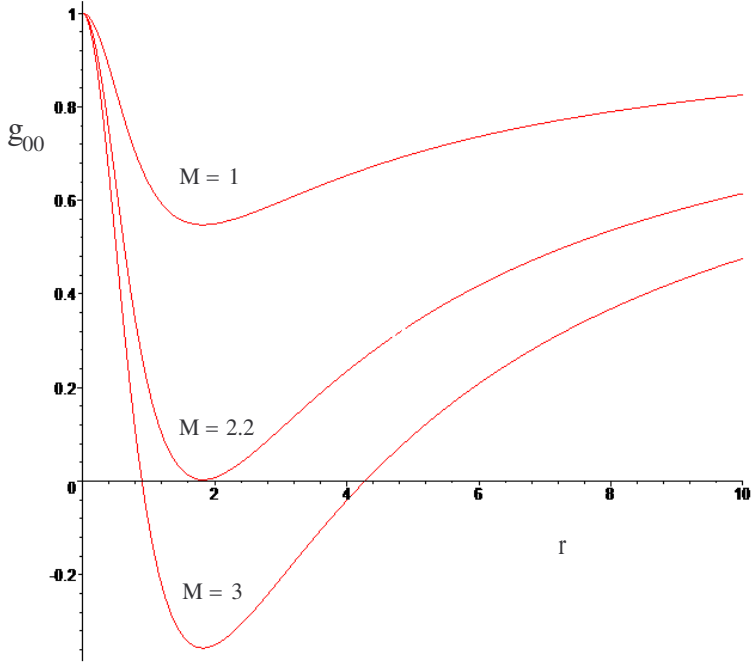}
\end{center}
\vspace{6.7 cm} \caption{\scriptsize {$g_{00}$ versus the radius $r$
for different values of mass $M$ utilizing Lorentzian smearing. We
set $\theta'=1$ (which is not exactly the same as $\theta=1$). This
figure is the same as the Fig. 1 with feasibility of having extremal
configuration with one degenerate event horizon when
$M=M_0\approx2.2$ ({\it i.e.}, the existence of a minimal non-zero
mass), and no event horizon when the mass of the black hole is
smaller than $M_0$. Also as figure again displays, the distance
between the horizons will grow by increasing the black hole mass
(two event horizons)in the same way as Gaussian profile. These
features show the lack of responsiveness of these consequences to
Gaussian formalism of the smearing.}}
\end{figure}
We set $\theta'=1$ (which is not equivalent to $\theta=1$).
Comparing these results with the results of Fig. 1 demonstrates the
close similarity of consequences in these two setup at least in
asymptotic values $r$. In both situations a minimum $g_{00}$ happens
at comparatively small\, $r$ \,with slightly comparable values of
$M$. The $M_0$ value is seen to be entirely similar in the two
situations. The preeminent distinction in the two approaches is
comprehended to take place around less than $M_0$ where there is the
mainly responsiveness to noncommutativity influences and the
detailed form of the matter distribution. However, we actually
should not have confidence to the details of our modeling when \,$r$
\,is excessively small. In fact, in the area where noncommutativity
effects precisely begin to be sensed, the detailed nature of the
sharpened mass distribution is not actually being explored. In a
recent paper [14], we have reported some outcomes about
extraordinary thermodynamical treatment for Planck-sized black hole
evaporation, {\it i.e.}, when $M$ is less than $M_0$. {\footnote
{Note that the foundation of these propositions possibly become as a
result of the fractal nature of spacetime at very short distances.
Theories such as $E$-infinity [15] and scale relativity [16] which
are on the basis of fractal structure of spacetime at very short
distances may prepare an appropriate framework to treat
thermodynamics of these very short distance systems.}} In this
manner, one encounters some uncommon thermodynamical features goes
to negative entropy, negative temperature and abnormal heat capacity
where the mass of the black hole becomes of the order of Planck mass
or less. It is also in this extreme situation that majority of the
distinctions between, {\it e.g.}, the Gaussian, Lorentzian or some
other forms of the smeared mass distribution would be anticipated to
commence to come into view. Therefore, we will henceforth use the
Gaussian-smeared mass distribution in our calculations just in the
circumstance that
$M\geq M_0$.\\

\section{Quantum Tunneling Near the Horizon}
We are now ready to discuss the quantum tunneling process in the
noncommutative framework. In accordance with Ref. [17], one can
express the general spherically symmetric line element in the form
\begin{equation}
ds^2=-[N_t(t,r)dt]^2+L(t,r)^2[dr+N_r(t,r)dt]^2+R(t,r)^2d\Omega^2.
\end{equation}
When we insert this expression into the Einstein-Hilbert action, due
to some restrictions and advantages {\it e.g.}, no time derivative
in the action, invariance of the action under reparametrization and
no singular behavior at the horizon, one finds
\begin{displaymath}
\left\{ \begin{array}{llll} N_t(t,r)=N_t(r)\\
N_r(t,r)=N_r(r)\\
L(t,r)=r\\
R(t,r)=1\\
\end{array} \right.
\end{displaymath}
To describe noncommutative quantum tunneling process where a
particle moves in dynamical geometry and pass through the horizon
without singularity on the path, we should use a coordinates system
that, unlike Schwarzschild coordinates, are not singular at the
horizon. These simply choices for $L$ and $R$ mentioned above (first
indicated by Painlev\'{e}{\footnote {Painlev\'{e} coordinates [18]
are especially convenient choices, which are obtained by definition
of a new noncommutative time coordinate, $
dt=dt_s+\frac{\sqrt{2M_{\theta}r}}{r-2M_{\theta}}dr $, where $t_s$
is the Schwarzschild time coordinate. Note that only the
Schwarzschild time coordinate is transformed. Both the radial
coordinate and angular coordinates remain the same.}}) can prepare
this purpose. Thus for a noncommutative Schwarzschild solution one
can easily acquire
\begin{displaymath}
\left\{ \begin{array}{ll} N_t(r)=1\\
N_r(r)=\sqrt{\frac{2M_\theta}{r}}\\
\end{array} \right.
\end{displaymath}
The noncommutative line element now immediately reads
\begin{equation}
ds^2 = -\bigg(1 - \frac{2M_\theta}{r} \bigg) dt^2 +
2\sqrt{\frac{2M_\theta}{r}} dtdr + dr^2 + r^2 d\Omega^2,
\end{equation}
The metric in these new coordinates is now stationary, non-static,
and there are neither coordinate nor intrinsic singularities (due to
noncommutativity). The equation of motion for a massless particle
(the radial null geodesic) is $\dot{r}\equiv\frac{dr}{dt}=\pm
N_t-N_r, $ where the upper sign (lower sign) corresponds to an
outgoing (ingoing) geodesic respectively. Since the horizon,
$r=r_H$, is concluded from the condition $N_t(r_H)-N_r(r_H)=0$, in
the vicinity of horizon $N_t-N_r$ treats as
\begin{equation}
N_t-N_r\simeq(r-r_H)\kappa(M)+O\left((r-r_H)^2\right),
\end{equation}
If we suppose that \,$t$\, increases towards the future, then the
above equations should be modified by the particle's
self-gravitation effect. Kraus and Wilczek [19] studied the motion
of particles in the $s$-wave as spherical massless shells in
dynamical geometry and developed self-gravitating shells in
Hamiltonian gravity. Further elaborations was performed by Parikh
and Wilczek [2]. On the other hand, Shankaranarayanan {\it et al}
have applied the tunneling approach to obtain the Hawking
temperature in different coordinates within a Complex paths scenario
[20]. This technique has been successfully applied to obtain a
global temperature for multi-horizon spacetimes [21]. In this paper,
we are going to develop Parikh-Wilczek method
to noncommutative coordinate coherent states.\\
We keep the total ADM mass ($M$) of the spacetime fixed, and allow
the hole mass fluctuated, due to the fact that we take into account
the response of the background geometry to an emitted quantum of
energy $E$ which moves in the geodesics of a spacetime with $M$
replaced by $M-E$. Thus we should replace $M$ by $M-E$ both in the equations (15) and (16).\\
Since the characteristic wavelength of the radiation is always
haphazardly small near the horizon due to the infinite blue-shift
there, the wave-number reaches infinity and therefore  WKB
approximation is reliable close to the horizon. In the WKB
approximation, the probability of tunneling or emission rate for the
classically forbidden region as a function of the imaginary part of
the particle's action at stationary phase would take the following
form
\begin{equation}
\Gamma\sim\exp(-2\textmd{Im}\, I).
\end{equation}
To calculate the imaginary part of the action we consider a
spherical shell consist of components massless particles each of
which travels on a radial null geodesic. We use these radial null
geodesics like an $s$-wave outgoing positive energy particle which
pass through the horizon outwards from $r_{in}$ to $r_{out}$ to
compute the $\textmd{Im}\, I$, as follows (on the condition that
$r_{in}>r_{out}$, where we should have:
$r_{in}=\frac{4M}{\sqrt{\pi}}\gamma\big(\frac{3}{2},\frac{r_{in}^2}{4\theta}\big)$
and
$r_{out}=\frac{4(M-E)}{\sqrt{\pi}}\gamma\big(\frac{3}{2},\frac{r_{out}^2}{4\theta}\big)$),
\begin{equation}
\textmd{Im}\,
I=\textmd{Im}\int_{r_{in}}^{r_{out}}p_rdr=\textmd{Im}\int_{r_{in}}^{r_{out}}\int_0^{p_r}dp'_rdr,
\end{equation}
one can alter the integral variable from momentum in favor of energy
by using Hamilton's equation $\dot{r}=\frac{dH}{dp_r}|_r$, where the
Hamiltonian is $H=M-E'$. We now evaluate the integral without
writing out the explicit form for the radial null geodesic. The $r$
integral can be done first by deforming the contour,
\begin{equation}\textmd{Im}\,
I=\textmd{Im}\int_{M}^{M-E}\int_{r_{in}}^{r_{out}}\frac{dr}{\dot{r}}dH=
-\textmd{Im}\int_{0}^{E}\int_{r_{in}}^{r_{out}}\frac{drdE'}{(r-r_H)\kappa(M-E')}
\end{equation}
The $r$ integral has a pole at the horizon which lies along the line
of integration and this yields ($-\pi i$) times the residue.
Therefore,
\begin{equation}
\textmd{Im}\,I=\pi\int_{0}^{E}\frac{dE'}{\kappa(M-E')},
\end{equation}
Here, reutilizing the first low of black hole thermodynamics,
$dM=\frac{\kappa}{2\pi}dS$, one can find the imaginary part of the
action as [22]
\begin{equation}
\textmd{Im}\,I=-\frac{1}{2}\int_{S_{NC}(M)}^{S_{NC}(M-E)}dS=-\frac{1}{2}\Delta
S_{NC}
\end{equation}
Hawking radiation as tunneling from the black hole event horizon was
also investigated in the context of string theory [22], and it was
exhibited that the emission rates in the high energy corresponds to
a difference between counting of states in the microcanonical and
canonical ensembles. In fact, the emission rates in the tunneling
approach just to first order in $E$, replace the Boltzmann factor in
the canonical ensemble $\Gamma\sim\exp(-\beta E)$, which is
described by the inverse temperature as the coefficient $\beta$. So,
the emission rates in the high energy are proportional to
$\exp(\Delta S)$ (see also [23]),
\begin{equation}
\Gamma\sim\exp(-2\textmd{Im}\,
I)\sim\frac{e^{S_{final}}}{e^{S_{initial}}}=\exp(\Delta
S)=\exp[S(M-E)-S(M)],
\end{equation}
where $\Delta S$ is the difference in black hole entropies before
and after emission. In other words, at higher energies the emission
probability depends on the final and initial number of microstates
available for the system. Thus, at higher energies the emission
spectrum cannot be precisely thermal due to the fact that the high
energy corrections arise from the physics of energy conservation
with noncommutativity corrections. In this model, one takes into
account the back-reaction results in a finite separation between the
initial and final radius as a result of self-gravitation effects of
outgoing shells that is the classically forbidden trajectory {\it
i.e.}, the barrier. On the other hand, according to energy
conservation the tunneling barrier is produced by a change in the
radius (the decreasing of the black hole
horizon) just by the emitted particle itself.\\
Let us now insert our result for noncommutative black hole entropy,
equation (9), into above equation and write the new
noncommutative-corrected tunneling probability as follows
$$
\Gamma\sim\exp(\Delta
S_{NC})=\exp[S_{NC}(M-E)-S_{NC}(M)]=\exp\Bigg(4\pi\bigg[(M-E)^2-\frac{3}{2}\theta
\bigg]\,{\cal{E}}\bigg(\frac{M-E}{\sqrt{\theta}}\bigg)+$$\begin{equation}12\sqrt{\pi\theta}(M-E)\exp
\bigg(-\frac{(M-E)^2}{\theta}\bigg)-4\pi\bigg[M^2-\frac{3}{2}\theta
\bigg]\,{\cal{E}}\bigg(\frac{M}{\sqrt{\theta}}\bigg)-12\sqrt{\pi\theta}M\exp\bigg(-\frac{M^2}{\theta}\bigg)\Bigg).
\end{equation}
It is simply observed that to linear order in $E$, two expressions
for $\Gamma$ in the microcanonical and canonical ensembles coincide.
So, manifestly the emission rate (23) deviates from the pure thermal
emission but is consistent with an underlying unitary quantum theory
[24]. We must note that the tunneling probability can also be
derived by writing out the explicit metric in the tunneling
computation, which would be leaded to the same result in spite of
more complicated calculations.\\

At this stage, we want to demonstrate that there are no correlations
between emitted particles even with the inclusion of the
noncommutativity corrections at least at late-times. (However, there
might be short-time correlations between the quanta emitted earlier
and the quanta emitted later on that decay to zero after the black
hole is equilibrated at late-times). This means it can be exhibited
that the probability of tunneling of two particles of energy $E_1$
and $E_2$ is precisely similar to the probability of tunneling of
one particle with their compound energies, $E=E_1+E_2$, {\it i.e.}
\begin{equation}
\Delta S_{E_1}+\Delta S_{E_2}=\Delta S_{(E_1+E_2)}\Rightarrow
\chi(E_1+E_2;E_1,E_2)=0,
\end{equation}
where the emission rate for a first quanta emitted, $E_1$, yields
$$
\Delta S_{E_1}=\ln\Gamma_{E_1}=4\pi\bigg[(M-E_1)^2-\frac{3}{2}\theta
\bigg]\,{\cal{E}}\bigg(\frac{M-E_1}{\sqrt{\theta}}\bigg)+12\sqrt{\pi\theta}(M-E_1)\exp
\bigg(-\frac{(M-E_1)^2}{\theta}\bigg)-$$\begin{equation}4\pi\bigg[M^2-\frac{3}{2}\theta
\bigg]\,{\cal{E}}\bigg(\frac{M}{\sqrt{\theta}}\bigg)-12\sqrt{\pi\theta}M\exp\bigg(-\frac{M^2}{\theta}\bigg),
\end{equation}
and similarly the emission rate for a second quanta emitted, $E_2$,
is given by
$$
\Delta
S_{E_2}=\ln\Gamma_{E_2}=4\pi\bigg[\Big((M-E_1)-E_2\Big)^2-\frac{3}{2}\theta
\bigg]\,{\cal{E}}\bigg(\frac{(M-E_1)-E_2}{\sqrt{\theta}}\bigg)+$$$$12\sqrt{\pi\theta}\Big((M-E_1)-E_2\Big)
\exp\Bigg(-\frac{\Big((M-E_1)-E_2\Big)^2}{\theta}\Bigg)-4\pi\bigg[(M-E_1)^2-\frac{3}{2}\theta
\bigg]\,{\cal{E}}\bigg(\frac{(M-E_1)}{\sqrt{\theta}}\bigg)-$$\begin{equation}
12\sqrt{\pi\theta}(M-E_1)\exp\bigg(-\frac{(M-E_1)^2}{\theta}\bigg).
\end{equation}
Finally, the emission rate for a single quanta emitted with the same
total energy, $E$, is given by
$$
\Delta
S_{(E_1+E_2)}=\ln\Gamma_{(E_1+E_2)}=4\pi\bigg[\Big(M-(E_1+E_2)\Big)^2-\frac{3}{2}\theta
\bigg]\,{\cal{E}}\bigg(\frac{M-(E_1+E_2)}{\sqrt{\theta}}\bigg)+$$
$$12\sqrt{\pi\theta}\Big(M-(E_1+E_2)\Big)\exp
\Bigg(-\frac{\Big(M-(E_1+E_2)\Big)^2}{\theta}\Bigg)-$$
\begin{equation}4\pi\bigg[M^2-\frac{3}{2}\theta
\bigg]\,{\cal{E}}\bigg(\frac{M}{\sqrt{\theta}}\bigg)-
12\sqrt{\pi\theta}M\exp\bigg(-\frac{M^2}{\theta}\bigg).
\end{equation}
It can be easily confirmed that these probabilities of emission are
uncorrelated. On the other hand, the statistical correlation
function, $\chi(E;E_1,E_2)$ is zero which leads to the independence
between different modes of radiation during the evaporation. Hence,
in this method the form of the corrections as back-reaction effects
even with inclusion of noncommutativity effects are not adequate by
themselves to retrieve information because there are no correlations
between different modes at least at late-times and information
doesn't come out with the Hawking radiation (for reviews on
resolving the so-called {\it information loss paradox}, see
[25-27]). Nevertheless, noncommutativity effect is adequate by
itself to preserve information due to the fact that in the
noncommutative framework black hole doesn't evaporate completely and
this leads to the existence of a minimal non-zero mass ({\it e.g.},
{\it a Planck-sized remnant containing the information}) which black
hole can reduce to it. So information might be preserved in this
remnant.

In string theory and loop quantum gravity, the entropy of black hole
has been achieved by direct microstate counting as follows (in units
of the Planck scale),
\begin{equation}
S_{QG}=4\pi M^2+\alpha\ln(16\pi M^2)+ O\bigg(\frac{1}{M^2}\bigg).
\end{equation}
It was recently suggested by the authors of Ref. [28] that the
Planck scale corrections to the black hole radiation spectrum via
tunneling can be written as
\begin{equation}
\Gamma\sim\exp(\Delta
S_{QG})=\exp\Big(S_{QG}(M-E)-S_{QG}(M)\Big)=\Bigg(1-\frac{E}{M}\Bigg)^{2\alpha}\exp\Bigg(-8\pi
ME\bigg(1-\frac{E}{2M}\bigg)\Bigg).
\end{equation}
Since, loop quantum gravity anticipates a negative value for
$\alpha$ (see {\it e.g.} [29]) which yields diverging emission rate
if $E\rightarrow M$, this leads to no suppressing the black hole
emission (although the suppression can only occur when $\alpha>0$,
which is not recommended at least by loop quantum gravity). But our
outcome is actually sensible, comparing the noncommutative result
for the emission rate (equation (23)) with the quantum gravity
result (equation (29)) shows that the noncommutative result is
reasonably successful in ceasing the black hole emission when
$(M-E)\rightarrow M_0$. In fact, the cases $(M-E)<M_0$ are the
noncommutativity-forbidden regions that the tunneling particle can
not be traversed through it. Therefore, the limit
$(M-E)\rightarrow0$ can not be applied by our process because of
existence of non-vanishing mass
at final phase of black hole evaporation.\\

\section{Summary and Remarks}
We summarize this paper with some remarks. In this paper,
generalization of the standard Hawking radiation via tunneling
through the event horizon based on the solution of the equation (17)
within the context of coordinate coherent state noncommutativity has
been studied and then the new corrections of the emission rate based
on spacetime noncommutativity has been achieved. In this study, we
see that there aren't any correlations between the tunneling rates
of different modes in the black hole radiation spectrum even in
noncommutative framework at least at late-times. In our opinion, if
we really believe the idea of stable black hole remnants due to the
fact that there are some exact continuous global symmetries in
nature [30], then we should accept that the information stays inside
the black hole and can be retained by a stable Planck-sized remnant.
In principle, there are four main outcomes of the black hole
evaporation:
\begin{itemize}
\item  The black hole can evaporate
completely, and information would disappear from our world.
\item  The black hole can completely disappear, but information emerges
in the final burst of radiation when the black hole shrinks to the
Planck size.
\item  There are correlations between different modes of radiation
during the evaporation that information appears continuously through
them.
\item  The black hole never disappears completely, and information
would preserve in a stable black hole remnant.
\end{itemize}
Indeed, it is not conceivable to date to give a clear answer to the
question of the black hole information paradox and this is
reasonable because there is no complete self-consistent quantum
theory of evaporating black holes (and gravity). In this paper we
have studied the reliability of the forth conjecture within a
noncommutative framework. We have shown that although there is no
correlation between the tunneling rates of different modes in the
black hole radiation spectrum in noncommutative spacetime at least
at late times, but noncommutativity has the capability to overcome
the information loss paradox via existence of stable black hole
remnant. At this stage we should stress that there is another point
of view on using relation (17). There is a problem here known as
"factor $2$ problem. Some authors such as Chowdhury [31] and Pilling
[32] have argued that relation (17) is not invariant under canonical
transformations but the same formula with a factor of $1/2$ in the
exponent is canonically invariant. As final remark we emphasize that
some authors have treated black hole thermodynamics in
noncommutative framework adapting a coordinate noncommutativity
against coherent state approach( see [33] and references therein). A
question then arises: what is the difference between these two
procedure? The standard way to handle noncommutative problems is
through the use of Wigner-Weyl-Moyal $\star$-product. That means to
use complex number commuting coordinates and shift non-commutativity
in the product between functions. This is mathematically correct,
but it is physically useless since any model written in terms of
star product, even the simplest field theory, becomes non-local and
it is not obvious how to handle non-local quantum field theory. One
proposed approach is perturbation in the $\theta$ parameter[34].
This is physically sensible since once expanded up to a given order
in $\theta$, the resulting field theory becomes local. The smeared
picture of particles based on coordinate coherent states defines
complex number coordinates as quantum mean values of the original
non-commuting ones between coordinate coherent states. In other
words, in this setup one can see commuting coordinates as classical
limit(in the quantum mechanical sense) of the non-commuting ones. In
this framework, the emergent semi-classical geometry keeps memory of
its origin. For example, free propagation of a point-like object is
described by a minimal width Gaussian wave-packet as has been
considered in our setup. So, the difference between two approaches
lies in the definition of quantum field
theoretical propagators.\\

Note added: After we have completed this work, Banerjee {\it et al}
have reported a similar treatment of the problem [35].


\begin{thebibliography}{10}
\bibitem{1}
S. W. Hawking, {\it Particle creation by black holes, Commun. Math.
Phys.} {\bf43}, 199 (1975).
\bibitem{2}
M. K. Parikh and F. Wilczek, {\it Hawking radiation as tunneling ,
Phys. Rev. Lett.} {\bf85}, 5042 (2000), [arXiv:hep-th/9907001].
\bibitem{3}
H. S. Snyder, {\it Quantized Spacetime, Phys. Rev.} {\bf71}, 38
(1947).
\bibitem{4}
N. Seiberg and E. Witten, {\it String theory and noncommutative
geometry, JHEP} {\bf9909}, 032 (1999).
\bibitem{5}
M. R. Douglas and N. A. Nekrasov, {\it Noncommutative field theory,
Rev. Mod. Phys.} {\bf73}, 977 (2001).
\bibitem{6}
A. Smailagic and E. Spallucci, {\it Feynman path integral on the
non-commutative plane, J. Phys. A} {\bf36}, L467 (2003),
[arXiv:hep-th/0307217].
\bibitem{7}
A. Smailagic and E. Spallucci, {\it UV divergence-free QFT on
noncommutative plane, J. Phys. A} {\bf36}, L517 (2003),
[arXiv:hep-th/0308193].
\bibitem{8}
A. Smailagic and E. Spallucci, {\it Lorentz invariance, unitarity
and UV-finiteness of QFT on noncommutative spacetime, J. Phys. A}
{\bf37}, 7169 (2004), [arXiv:hep-th/0406174].
\bibitem{9}
P. Nicolini, A. Smailagic and E. Spallucci, {\it Noncommutative
geometry inspired Schwarzschild black hole, Phys. Lett. B }
{\bf632}, 547 (2006), [arXiv:gr-qc/0510112].
\bibitem{10}
X. Calmet and A. Kobakhidze, {\it Noncommutative General Relativity,
Phys. Rev. D} {\bf72}, 045010 (2005), [arXiv:hep-th/0506157]; X.
Calmet and A. Kobakhidze, {\it Second Order Noncommutative
Corrections to Gravity, Phys. Rev. D} {\bf74}, 047702 (2006),
[arXiv:hep-th/0605275]; A. H. Chamseddine, {\it Deforming Einstein's
Gravity, Phys. Lett. B} {\bf504}, 33 (2001), [arXiv:hep-th/0009153];
P. Aschieri {\it et al}, {\it A Gravity Theory on Noncommutative
Spaces, Class. Quant. Grav.} {\bf22}, 3511 (2005),
[arXiv:hep-th/0504183].
\bibitem{11}
J. E. Moyal, {\it Quantum mechanics as a statistical theory, Proc.
Cambridge Phil. Soc.} {\bf45} (1949) 99; M. S. Barlett and J. E.
Moyal, {\it The exact transition probabilities of quantum-mechanical
oscillators calculated by the phase-space method, Proc. Cambridge
Philos.} {\it45} (1949) 545.
\bibitem{12}
P. Nicolini, {\it Vacuum energy momentum tensor in (2+1) NC scalar
field theory}, [arXiv:hep-th/0401204] ; A. Gruppuso, {\it Newton's
law in an effective non-commutative space-time, J. Phys. A} {\bf38},
2039 (2005), [arXiv:hep-th/0502144].
\bibitem{13}
T. G. Rizzo, {\it Noncommutative inspired black holes in extra
dimensions, JHEP} {\bf0609}, 021 (2006), [arXiv:hep-ph/0606051].
\bibitem{14}
K. Nozari and S. H. Mehdipour, {\it Failure of Standard
Thermodynamics in Planck Size Black Hole System, Chaos, Solitons and
Fractals} {\bf32}, 1 (2007), [arXiv:hep-th/0610076].
\bibitem{15}
M. S. El Naschie, {\it The concepts of E-infinity: An elementary
introduction to the Cantorian-fractal theory of quantum physics,
Chaos, Solitons and Fractals} {\bf 22}, 495 (2004).
\bibitem{16}
L. Nottale, {\it Fractal space-time and microphysics: towards a
theory of scale relativity}, World Scientific, Singapore, 1993.
\bibitem{17}
P. Kraus and F. Wilczek, {\it Some Applications of a Simple
Stationary Line Element for the Schwarzschild Geometry, Mod. Phys.
Lett. A} {\bf40}, 3713 (1994), [arXiv:gr-qc/9406042].
\bibitem{18}
P. Painlev\'{e}, {\it La m\'{e}canique classique et la th\'{e}orie
de la relativit\'{e}, Compt. Rend. Acad. Sci. (Paris)} {\bf173}, 677
(1921).
\bibitem{19}
P. Kraus and F. Wilczek, {\it Self-Interaction Correction to Black
Hole Radiance, Nucl. Phys. B} {\bf433}, 403 (1995),
[arXiv:gr-qc/9408003]; P. Kraus and F. Wilczek, {\it Some
Applications of a Simple Stationary Line Element for the
Schwarzschild Geometry, Mod. Phys. Lett. A} {\bf9}, 3713 (1994),
[arXiv:gr-qc/9406042].
\bibitem{20}
S. Shankaranarayanan, T. Padmanabhan and K. Srinivasan, {\it Hawking
radiation in different coordinate settings: Complex paths approach,
Class. Quant. Grav.} {\bf 19}, 2671 (2002), [arXiv:gr-qc/0010042];
S. Shankaranarayanan, K. Srinivasan and T. Padmanabhan, {\it Method
of complex paths and general covariance of Hawking radiation , Mod.
Phys. Letts. A} {\bf16}, 571 (2001), [arXiv:gr-qc/0007022].
\bibitem{21}
S. Shankaranarayanan,  {\it Temperature and Entropy of
Schwarzschild-de Sitter space-time, Phys. Rev. D} {\bf67} 084026
(2003), [arXiv:gr-qc/0301090].
\bibitem{22}
E. Keski-Vakkuri and P. Kraus, {\it Microcanonical D-branes and Back
Reaction, Nucl. Phys. B} {\bf491}, 249 (1997),
[arXiv:hep-th/9610045].
\bibitem{23}
S. Massar and R. Parentani, {\it How the Change in Horizon Area
Drives Black Hole Evaporation, Nucl. Phys. B} {\bf575}, 333 (2000),
[arXiv:gr-qc/9903027].
\bibitem{24}
M. K. Parikh, {\it A Secret Tunnel Through The Horizon, Int. J. Mod.
Phys. D} {\bf13}, 2351 (2004), [arXiv:hep-th/0405160]; M. K. Parikh,
{\it Energy Conservation and Hawking Radiation},
[arXiv:hep-th/0402166].
\bibitem{25}
J. Preskill, {\it Do Black Holes Destroy Information?}, Publication:
{\it An international symposium on Black Holes, Membranes, Wormholes
and Superstrings, Houston Advanced Research Center, 16-18 January
1992. Edited by Sunny Kalara and D. V. Nanopoulos. Singapore: World
Scientific, 1993, p.22}, [arXiv:hep-th/9209058].
\bibitem{26}
D. N. Page, {\it Information in black hole radiation, Phys. Rev.
Lett.} {\bf71}, 3743 (1993), [arXiv:hep-th/9306083].
\bibitem{27}
J. G. Russo, {\it The Information Problem in Black Hole Evaporation:
Old and Recent Results}, [arXiv:hep-th/0501132].
\bibitem{28}
A. J. M. Medved and E. Vagenas, {\it On Hawking Radiation as
Tunneling with Logarithmic Corrections, Mod. Phys. Lett. A} {\bf20},
1723 (2005), [arXiv:gr-qc/0505015]; M. Arzano, A. Medved and E.
Vagenas, {\it Hawking Radiation as Tunneling through the Quantum
Horizon, JHEP} {\bf0509}, 037 (2005), [arXiv:hep-th/0505266].
\bibitem{29}
K. A. Meissner, {\it Black hole entropy in Loop Quantum Gravity,
Class. Quant. Grav.} {\bf21}, 5245 (2004), [arXiv:gr-qc/0407052].
\bibitem{30}
J. D. Bekenstein, {\it Nonexistence of baryon number for static
black holes, Phys. Rev. D} {\bf5}, 1239 (1972).
\bibitem{31}
Borun D. Chowdhury, {\it Problems with Tunneling of Thin Shells from
Black Holes, Pramana} {\bf70}, 593 (2008), [arXiv:hep-th/0605197].
\bibitem{32}
T. Pilling, {\it Black hole thermodynamics and the factor of 2
problem, Phys. Lett. B} {\bf660}, 402 (2008),
[arXiv:gr-qc/0709.1624].
\bibitem{33}
K. Nozari and B. Fazlpour, {\it Thermodynamics of Noncommutative
Schwarzschild Black Holes,  Mod. Phys. Lett. A } {\bf 22}, 2917
(2007), [arXiv:hep-th/0605109].
\bibitem{34}
M. Chaichian, M. M. Sheikh-Jabbari and A. Tureanu, {\it Hydrogen
Atom  Spectrum and the Lamb Shift in Noncommutative QED,
Phys.Rev.Lett.} {\bf 86}, 2716  (2001),  [arXiv:hep-th/0010175].
\bibitem{35}
R. Banerjee, B. Ranjan Majhi and S. Samanta, {\it Noncommutative
Black Hole Thermodynamics}, [arXiv:0801.3583].

\end{thebibliography}
\end{document}